\documentclass[fleqn,12pt,twoside]{article}
\usepackage{espcrc1}
\usepackage{amssymb}
\usepackage{amsmath}

\usepackage{graphicx}

\title{Quark matter in Neutron Star Mergers}

\author{R. Oechslin\address[BSL]{Departement f\"ur Physik und Astronomie der Universit\"at Basel, Klingelbergstrasse 82, 4056 Basel, Switzerland} 
        \thanks{Email: roland.oechslin@unibas.ch},
        G. Poghosyan\addressmark[BSL],
        K. Ury\=u
\address{Department of Physics, University of Wisconsin-Milwaukee, 
Milwaukee, WI 53211, USA}
\address{SISSA, via Beirut 2/4, 34014 Trieste, Italy}}

\begin{document}

\maketitle

\section{Introduction}

Binary neutron star mergers are expected to be one of the most promising source of gravitational waves (GW) for the network of laser interferometric and bar detectors becoming operational in the next few years. The merger wave signal is expected to be sensitive to the interior structure of the neutron star (NS). The structure of high density phases of matter is under current experimental investigation in heavy-ion collisions \cite{hallmann}. We investigate the dependence of the merger process and its GW signal on the presence of quarks in these phases by performing numerical simulations, where the smoothed particle hydrodynamics (SPH) method and 
the conformally flat approximation for the 3-geometry in general 
relativistic gravity are implemented \cite{roland}.

\section{Equation of state}

\begin{figure}
\begin{minipage}[t]{76mm}
\includegraphics[width=75mm]{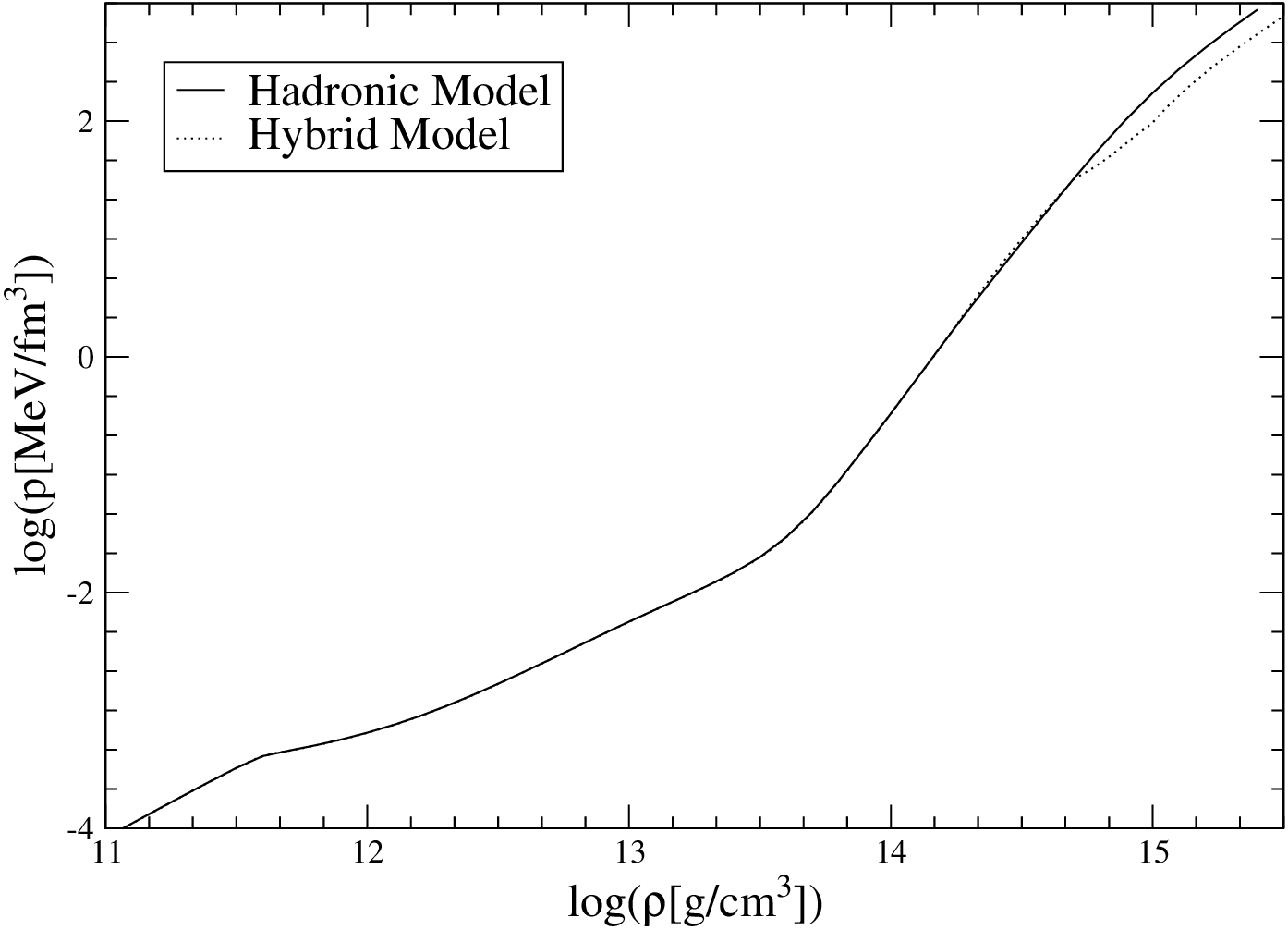}
\caption{Pressure-density relation for the two EoSs. The two equations coincide up to $\rho\simeq5\times10^{14}$ where the fluid enters the quark hadron mixed phase in the hybrid case. Above $\rho\simeq10^{15}$ only a pure quark phase exists. Note the low adiabatic index of the hybrid model EoS in the mixed phase of about 1.5.}
\label{fig:rhop}
\end{minipage}
\hspace{\fill}
\begin{minipage}[t]{79mm}
\includegraphics[width=77mm]{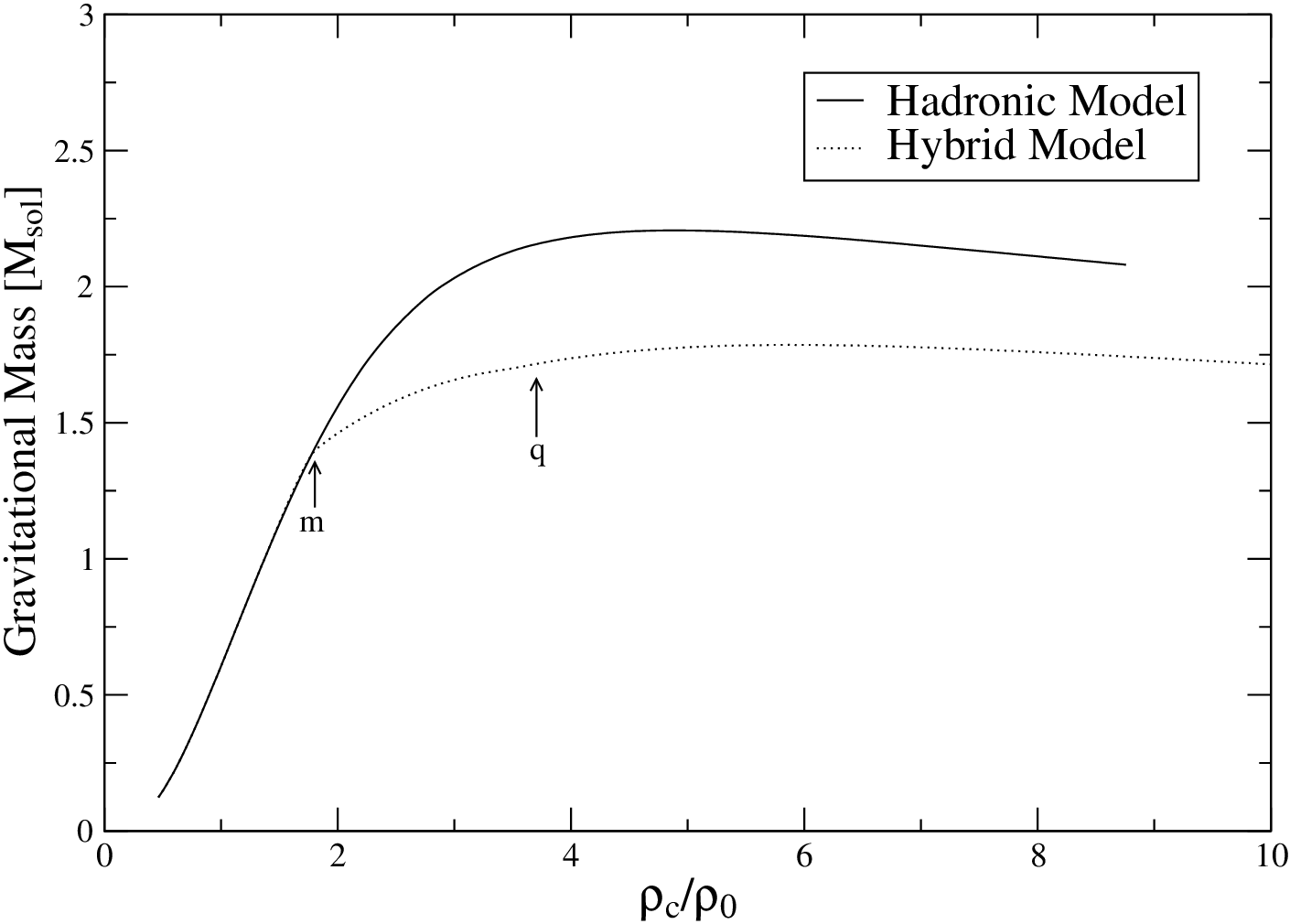}
\caption{Central density-mass relation of a series of spherical static NS using both EoSs. The maximum mass for the hadronic EoS is $M\simeq2.21M_\odot$ while it is $M\simeq1.78M_\odot$ in the hybrid case. Stars above 'm' have a quark-hadron mixed core in the interior, above 'q' a pure quark core.}
\label{fig:rhocM}
\end{minipage}
\end{figure}

Two types of EoS for high density matter, one with quarks and the other without, are used in our simulations. In the hadronic model, we describe the stellar matter in the high-density region using a EoS based on the relativistic mean field theory with the TM1 parameter set using a Lagrangian which contains the low-lying baryons and mesons as relativistic fields. The matter at low densities containing a small number of protons embedded in the neutron sea with electrons, muons, $\alpha$-particles and heavy nuclei is described using a Thomas-Fermi approximation \cite{shen}. In the hybrid model, we switch to a MIT-bag model based EoS (B=90MeV/fm${}^3$) to describe the exotic phases of superdense matter with up and down quarks and a negligible small amount of electrons and muons \cite{kettner}. For the possible mixed phase of hadronic and quark matter which appears for densities between 1.8$\rho_0$ ($\rho_0$:nuclear saturation density) and 3.7$\rho_0$, we use the multiconserved charge phase transition construction \cite{glendenning}. In both models, T=0 and $\beta$-equilibrium is assumed (see Fig. \ref{fig:rhop}).

\section{Results}
We start both simulations from irrotational quasi-equilibrium configurations \cite{uryu} with identical stars each having a gravitational mass in isolation of $M=1.5M_\odot$ \cite{chubarian}. In the case of the hybrid model, 
the component stars have mixed phase matter cores with a radius of 
about 2km. The fluid in the mixed phase has a very low adiabatic index 
between 1 and 1.5. Therefore, for constant restmass, a spherical star has a larger central density in the hybrid model than in the hadronic model.

\begin{figure}
\begin{minipage}[t]{76mm}
\includegraphics[width=75mm]{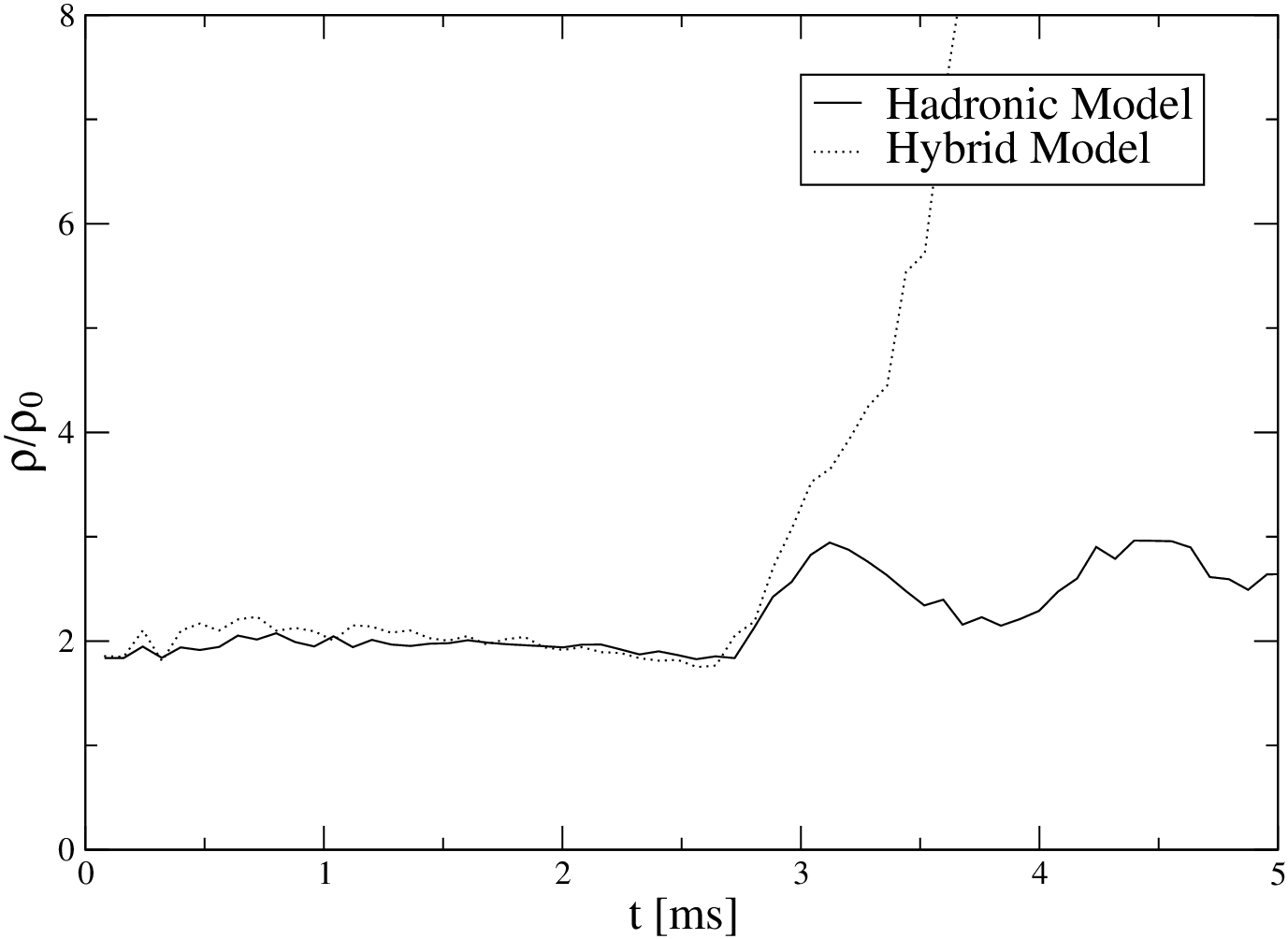}
\caption{The maximum density in units of $\rho_0$. Note the difference 
during the early inspiral phase and the collapse phase.}
\label{fig:cdens}
\end{minipage}
\hspace{\fill}
\begin{minipage}[t]{79mm}
\includegraphics[width=77mm]{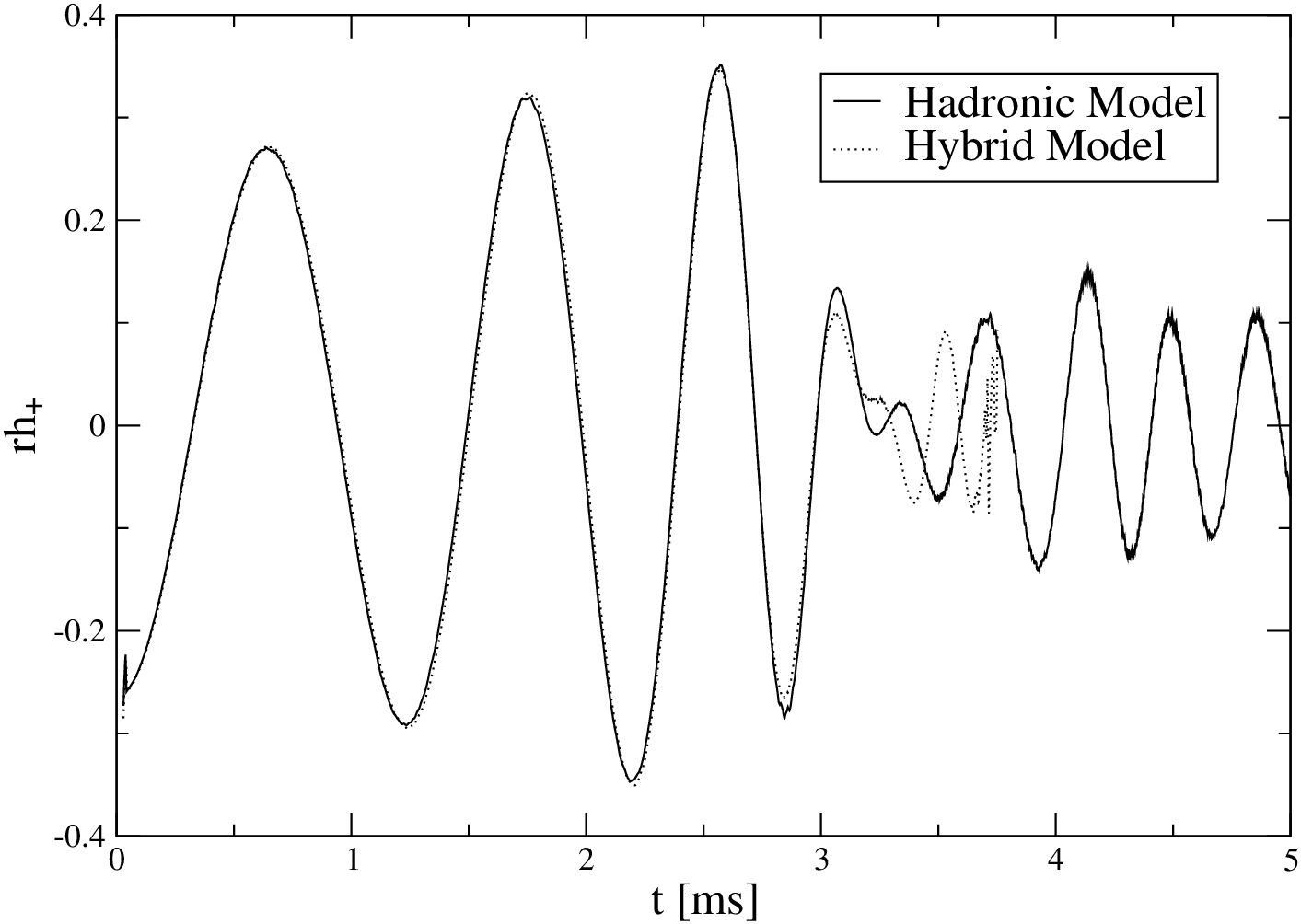}
\caption{Gravitational waveforms of both models. The inspiral part is almost 
identical. The collapse of the hybrid model results in a faster 
oscillation.}
\label{fig:gravsign}
\end{minipage}
\end{figure}

In Fig.~\ref{fig:cdens} we plot the evolution of the maximum density.  
As expected, the value of the hybrid model around the initial phase is 
slightly larger due to a certain amount of mixed phase material.  
At $t\sim2.5$ms, this difference vanishes because the maximum density 
drops below the phase transition density of $1.8\rho_0$ as a result of 
tidal deformation. Soon after, the two stars start merging 
into a single object. The evolution of the merger objects largely depends
on the EoS. In the hybrid model, the merger object is contracted further 
to form a high density core because of infalling matter and appearing 
quark matter which forms a compact mixed phase core.  
Finally the gravitation of the high density core overcome the pressure and the centrifugal forces to balance with it and the core starts to collapse. 
On the other hand, the hadronic EoS is much stiffer and the matter is harder to compress. Hence, the gravitational contraction is turned into an 
oscillation of the central object.

The GW signals (Fig.~\ref{fig:gravsign}) are almost identical until 
their amplitudes reach the maximum values.  Around $t\sim3$ms, 
the amplitude drops down to the minimum as the two component stars merge.  
In the hybrid model, since the merger object contracts faster, 
the quadrupole moment becomes smaller and the angular velocity becomes 
larger compared to that of hadronic model. 
This is visible in the wave signal as a smaller amplitude and a phase shift. 
A moment later, when the merger object begins to collapse, the signals begin 
to deviate completely from each other. Due to the higher central density, the hybrid model merger object rotates and oscillates faster which results in a higher frequency of the GW signal.

The two models can be distinguished by the GW signal from the merger phase 
($t\gtrsim2.5$ms) whose frequency is above $\sim1$kHz.
However, in this range, the broadband interferometers fail due to shot noise and narrowband resonant bar detectors are needed for a successful detection. We are expecting that the merger process of binaries with larger masses containing pure quark cores ($M\gtrsim1.7M_\odot$, see Fig. \ref{fig:rhocM}) will also allow for a distinction of the models by the inspiral part of the GW signal which is measurable with the broadband interferometer detectors.


\begin{thebibliography}{9}
\bibitem{hallmann}T. J. Hallmann et. al. (eds.), {\it Quark Matter 2001: Proceedings}, Nucl. Phys. 698 (2002).
\bibitem{roland} R. Oechslin, S. Rosswog, F. K. Thielemann, Phys. Rev. D 65 (2002) 103005. 
\bibitem{shen} H. Shen, H. Toki, K. Oyamatsu, K. Sumiyoshi, Prog. Theor. Phys. 100 (1998) 1013.
\bibitem{kettner}Ch. Kettner et al., Phys. Rev. D 51 (1995) 1440.
\bibitem{glendenning} N. K. Glendenning, Phys. Rev. D 46 (1992) 1274.
\bibitem{uryu}K. Ury\=u, Y. Eriguchi, Phys. Rev. D 61 (2000) 124023.
\bibitem{chubarian}E. Chubarian, H. Grigorian, G. Poghosyan, D. Blaschke, Astron. Astroph. 357 (2000) 968.
\end{thebibliography}
\end{document}